\documentclass
[twocolumn] 
{aa}                             

\usepackage[T1]{fontenc}
\usepackage{graphicx}
\usepackage{bm}                     
\usepackage{amssymb}
\usepackage[varg]{txfonts}           
\usepackage{natbib} \bibpunct{(}{)}{;}{a}{}{,} 

\def\bea{\begin{eqnarray}}
\def\eea{\end{eqnarray}}
\def\be{\begin{equation}}
\def\ee{\end{equation}}
\def\ms{M_\odot}

\begin{document}

\title{Structure of the hadron-quark mixed phase in protoneutron stars}


\author{
H. Chen,$^{1,2}$
G. F. Burgio,$^1$ H.-J. Schulze,$^{1,3}$ and N. Yasutake$^4$}

\institute{
$^1$ INFN Sezione di Catania, Dipartimento di Fisica,
Universit\'a di Catania, Via Santa Sofia 64, 95123 Catania, Italy\\
$^2$ Physics Department, China University of Geoscience,
Wuhan 430074, China\\
$^3$ Yukawa Institute for Theoretical Physics, Kyoto University,
Kyoto 606-8502, Japan\\
$^4$ Department of Physics, Chiba Institute of Technology,
2-1-1 Shibazono, Narashino, Chiba 275-0023, Japan
}

\date{\today}

\abstract{
We study the hadron-quark phase transition
in the interior of hot protoneutron stars,
combining the Brueckner-Hartree-Fock approach for hadronic matter with
the MIT bag model or the Dyson-Schwinger model for quark matter.
We examine the structure of the mixed phase constructed
according to different prescriptions for the phase transition,
and the resulting consequences for stellar properties.
We find important effects for the internal composition,
but only very small influence on the global stellar properties.
}


\authorrunning{H. Chen et al.} 
\titlerunning{Structure of the hadron-quark mixed phase in protoneutron stars}


\maketitle


\section{Introduction}

A protoneutron star (PNS) is formed after
the gravitational collapse of the core of a massive star
($M \gtrsim 8\ms$),
exploding in a type-II supernova \citep{shapiro,bethe}.
Although the explosion mechanism is still not fully explained \citep{burrows12},
some general features can be considered as robust.
In fact, just after the core bounce,
the PNS is very hot and lepton-rich,
and neutrinos are trapped for a few seconds.
The following evolution of the PNS is dominated by neutrino diffusion,
causing deleptonization and subsequently cooling.
Ultimately, the neutron star (NS) achieves thermal equilibrium
and stabilizes at practically zero temperature without trapped neutrinos.

The theoretical description of the formation of a PNS
requires an accurate treatment of the microphysics of the collapsing matter,
in particular of neutrino transport and related processes
\citep{burrows,rep,burrows12}.
Moreover the violent dynamical processes occurring in the
contracting-exploding star need to be treated in 
the framework of general relativity [see \citet{ott} for a recent review].  
The physical processes which contribute to the subsequent PNS evolution,
such as nuclear and weak interactions and energy and lepton number
transport by neutrino diffusion,
are very difficult to include in dynamical simulations. 
Thus, most simulations of the gravitational core collapse to a PNS
end shortly after the core bounce and the launch of the supernova explosion,
typically after a few hundreds of milliseconds, 
and only a few dynamical simulations extend to the first minute
of the PNS life \citep{ponsevo,fischer}.

During the evolution of a PNS into a NS, a hadron-quark (HQ)
phase transition could take place in the central region of the
star \citep{cooke,lugon,ponsevo,pons,steiner,nemes,protomit,pnsymit},
and this would alter substantially the composition of the core.  
In fact the heaviest NS,
close to the maximum mass (about two solar masses),
are characterized by central baryon densities larger than
$1/{\rm fm}^{3}$,
as predicted by calculations   
based on a microscopic nucleonic equation of state (EOS).

The study of hybrid stars is also important from another point of view:
Purely nucleonic EOS are able to accommodate fairly large (P)NS maximum masses
\citep{bbb,akma,gle,zhou,zhli},
but the appearance of hyperons in beta-stable matter could
strongly reduce this value \citep{gle,mmy,zhli,carroll,djapo,nsesc}.
In this case the presence of non-baryonic, i.e., ``quark'' matter
would be a possible manner to stiffen the EOS
and reach larger NS masses \citep{nsquark,nscdm,romat,weiss}.
Heavy NS thus would be hybrid quark stars.

In previous articles \citep{proto,isen1,isen2,ourgw}
we have studied static properties of PNS
using a finite-temperature hadronic EOS including also hyperons \citep{pnsy}
derived within the Brueckner-Bethe-Goldstone theory
of nuclear matter \citep{bhf}.
An eventual HQ phase transition was modeled within an extended
MIT bag model \citep{protomit,pnsymit}
or a more sophisticated quark model,
the Dyson-Schwinger model (DSM) \citep{dsm1,dsm4,dsm2,dsm3,ourdsm1,ourdsm2}.

The purpose of the present work is to complement our previous articles
by studying details of the HQ phase transition occuring
in hybrid stars and their implications for the structure of a PNS,
in particular the question whether global (P)NS observables are sensitive to
and thus may reveal information on
the internal stellar structure.

In Sec.~II we briefly sketch
the theoretical approaches which we use for modeling the hadron and the
quark phases, and in Sec.~III we describe the corresponding pure phases.
The structure of the mixed phase is discussed in Sec.~IV, and the results
are illustrated in Sec.~V. 
Finally, in Sec.~VI we summarize our conclusions.

\section{Equations of state}
\label{s:eos}

The EOSs for hadronic matter (HM) and quark matter (QM) that we use
in this work, have been amply discussed in previous publications
\citep{nsquark,nscdm,protomit,proto,isen1,isen2,pnsy,ourdsm1,ourdsm2},
where all necessary details can be found.
Our hadronic EOS is obtained from Brueckner-Hartree-Fock (BHF) calculations of
(hyper)nuclear matter \citep{hypmat1,hypmat2,hypmat3}
based on realistic potentials
[the Argonne $V_{18}$ nucleon-nucleon \citep{v18}
and the Nijmegen NSC89 nucleon-hyperon \citep{nsc89} in this case]
supplemented by nucleonic Urbana UIX three-body forces \citep{uix1,uix2,uix3},
and extended to finite temperature \citep{pnsy}.
We employ two different representative models for QM,
an extended MIT bag model
[the model with 
a density-dependent bag constant
of \citet{nsquark,nscdm,protomit}]
and a Dyson-Schwinger model 
[the model DS4 of \citet{ourdsm1,ourdsm2}],
which yield in fact quite different internal structures of hybrid stars.

Those theoretical calculations provide the free energy density of
the bulk system (pure HM or QM)
as a function of the relevant partial number densities $n_i$
and the temperature, $f(\{n_i\},T)$,
from which all thermodynamic quantities of interest can be computed,
namely, the chemical potentials $\mu_i$, pressure $p$,
entropy density $s$, and internal energy density $\varepsilon$ read as
\bea
 \mu_i &=& {{\partial f}\over{\partial n_i}} \:,
\\
 p &=& n_B^2 {{\partial (f/n_B)}\over{\partial n_B} }
 = \sum_i \mu_i n_i - f  \:,
\label{e:p}
\\
 s &=& -{{\partial f}\over{\partial T}} \:,
\\
 \varepsilon &=& f + Ts \:,
\label{e:eps}
\eea
where $n_B$ is the total baryon number density.
These quantities allow to determine the stellar matter composition
and the EOS,
which is the fundamental input for solving the Tolman-Oppenheimer-Volkoff
equations of (P)NS structure.

\section{Pure phases}
\label{s:pur}

In neutrino-trapped beta-stable (hyper)nuclear or quark matter
the chemical potential $\mu_i$ of any particle 
$i=n,p,\Lambda,\Sigma^-,u,d,s,e,\mu,\nu_e,\nu_\mu,\ldots$
is uniquely determined by the conserved quantities
baryon number $B_i$, electric charge $C_i$,
and weak charges (lepton numbers) $L^{(l)}_i,\; l=e,\mu$
with the corresponding set of independent chemical potentials
$\mu_B,\,\mu_C,\,\mu_{L^{(e)}},\,\mu_{L^{(\mu)}}$:
\be
 \mu_i = B_i\mu_B + C_i\mu_C + L^{(e)}_i \mu_{L^{(e)}}
 + L^{(\mu)}_i \mu_{L^{(\mu)}} \:.
\label{e:mui}
\ee
In this work we neglect muons and muon neutrinos due to their low fractions
and negligible impact on global stellar properties,
hence use simply $\nu\equiv\nu_e$, $L\equiv L^{(e)}$.
The relations between chemical potentials and partial densities
for hadrons and quarks
are given by the microscopic models mentioned before, 
while leptons are treated as free fermions.
With such relations, the bulk system in each phase can be solved
for a given baryon density,
imposing the charge neutrality condition
and lepton number conservation:
\bea
 n_B &=& \sum_i n_i B_i\:,
\\
 0 &=& \sum_i n_i C_i\:,
\label{e:neutral}
\\
 Y_e n_B &=& \sum_i n_i L_i^{(e)} \:.
\label{e:lepfrac}
\eea
When the neutrinos $\nu_e$ are untrapped,
the lepton number is not conserved any more,
the density and the chemical potential of $\nu_e$ vanish,
and the above equations simplify accordingly.

\section{Mixed phase constructions}
\label{s:mix}

We are interested in the HQ phase transition in PNS
and consider therefore the usual oversimplified standard conditions, 
namely trapped hot matter with a fixed lepton fraction
$Y_e \equiv (n_e + n_\nu)/n_B = 0.4$ 
and either an isentropic, $S/A=2$,
or an isothermal, $T=40$ MeV, temperature profile.
One could consider more realistic profiles \citep{ourgw},
but we focus in this work
on the difference between phase transition constructions.

A fully microscopic treatment of the HQ mixed phase involving
finite-size (pasta) structures can only be performed numerically 
\citep{mixtat1,mixtat2,mixtat3,maru,yasu,yasulet,yasurev}.
One introduces Coulomb energies and surface energies via a HQ surface tension
and then minimizes the (free) energy of a Wigner-Seitz (WS) cell,
allowing for different geometrical structures
of the quark phase embedded in the hadron phase and vice versa.
The output are the optimal size and geometry of the cell,
as well as the local distributions of the individual particle species,
and also the Coulomb field inside the cell.
Some illustrative examples can be found in the given references.

This is a very time-consuming and not very transparent numerical procedure.
It is therefore convenient to search for reliable approximations
to this procedure, and in this article we compare two prescriptions 
corresponding to two limiting cases of the full numerical procedure,
that are termed global charge neutral (GCN) and local charge neutral (LCN)
mixed phase.

The first procedure is well known as Bulk Gibbs or Glendenning construction
\citep{glen1,glen2} from the zero-temperature case
and corresponds to a ``small'' WS cell
[compared to the electromagnetic Debye screening length, which is about
5-10 fm \citep{mix1,mix2,mix3,mix4,mix5}]
caused by a ``small'' HQ surface tension.
In this case the electromagnetic potential is practically constant
throughout the cell, and an electric field does not exist.
Consequently the electron density is also constant,
while the hadron and quark densities and their electric charges
are different in order to fulfill the conditions of pressure and baryon chemical
potential equality at the HQ interface.
In the case of neutrino trapping,
the neutrino densities have also to be equal in both phases,
\be
 n_\nu^H = n_\nu^Q \:, 
\label{e:nnu}
\ee
which together with the equal electron densities implies
equal lepton densities $n_L=n_e+n_\nu$
(but not lepton fractions $Y_e=n_L/n_B$) in both phases.
Altogether we have therefore 
the equality of the intensive thermodynamical quantities in both phases:
\bea
 \mu_B^{H} &=& \mu_B^{Q} \:,
\\
 \mu_C^{H} &=& \mu_C^{Q} \:, 
\label{e:muc}
\\
 \mu_L^{H} &=& \mu_L^{Q} \:,
\\
 p_H &=& p_Q \:,
\\
 T_H &=& T_Q \:,
\eea
which together with the general rule Eq.~(\ref{e:mui})
determines the composition of the system
for given overall baryon density $n_B$,
vanishing electric charge,
fixed lepton fraction $Y_e$ in the trapped case,
and eventually a prescribed entropy profile $S/A(n_B)$:
\bea
 (1 - \chi)n_B^H + \chi n_B^Q &=& n_B \:,
\\
 (1 - \chi)n_C^H + \chi n_C^Q &=& 0 \:,
\\
 (1 - \chi)n_L^H + \chi n_L^Q &=& n_B Y_e \:,
\label{e:nl}
\\
 (1 - \chi)s^H   + \chi s^Q   &=& n_B S/A \:,
\eea
where $\chi$ is the volume fraction occupied by the quark phase
and the last equation determines the local temperature.


\begin{figure}[t]
\includegraphics[scale=0.38,bb=60 90 824 860]{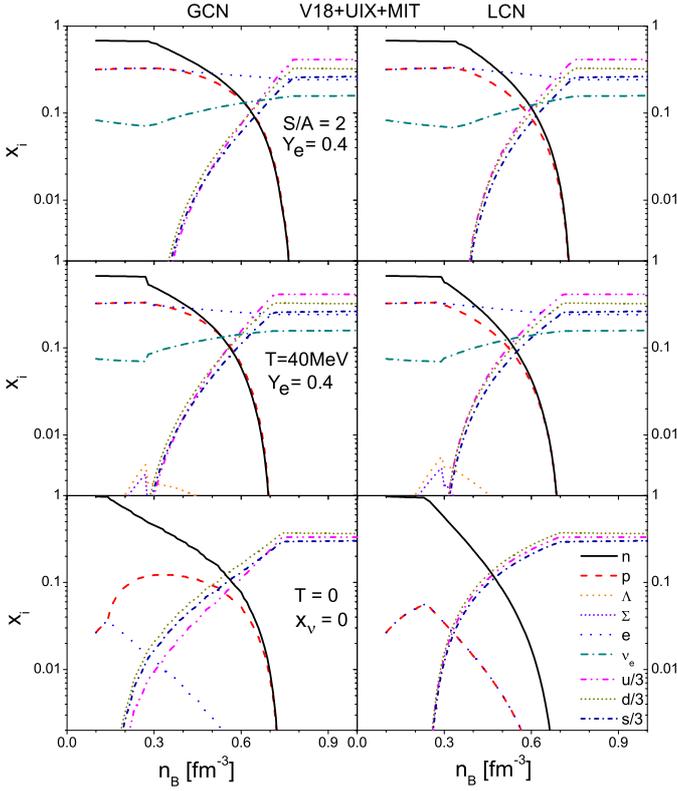}
\caption{
Relative populations $x_i=n_i/n_B$ for different stellar 
compositions: 
trapped matter with $S/A=2$ (upper panels) or $T=40$ MeV (central panels),
and untrapped matter with $T=0$ (lower panels). 
The GCN (left panels) or LCN (right panels) construction for the
mixed phase is employed together with the MIT bag model for the quark phase.}
\label{f:xi_mit}
\end{figure}

\begin{figure}[t]
\includegraphics[scale=0.38,bb=60 90 824 860]{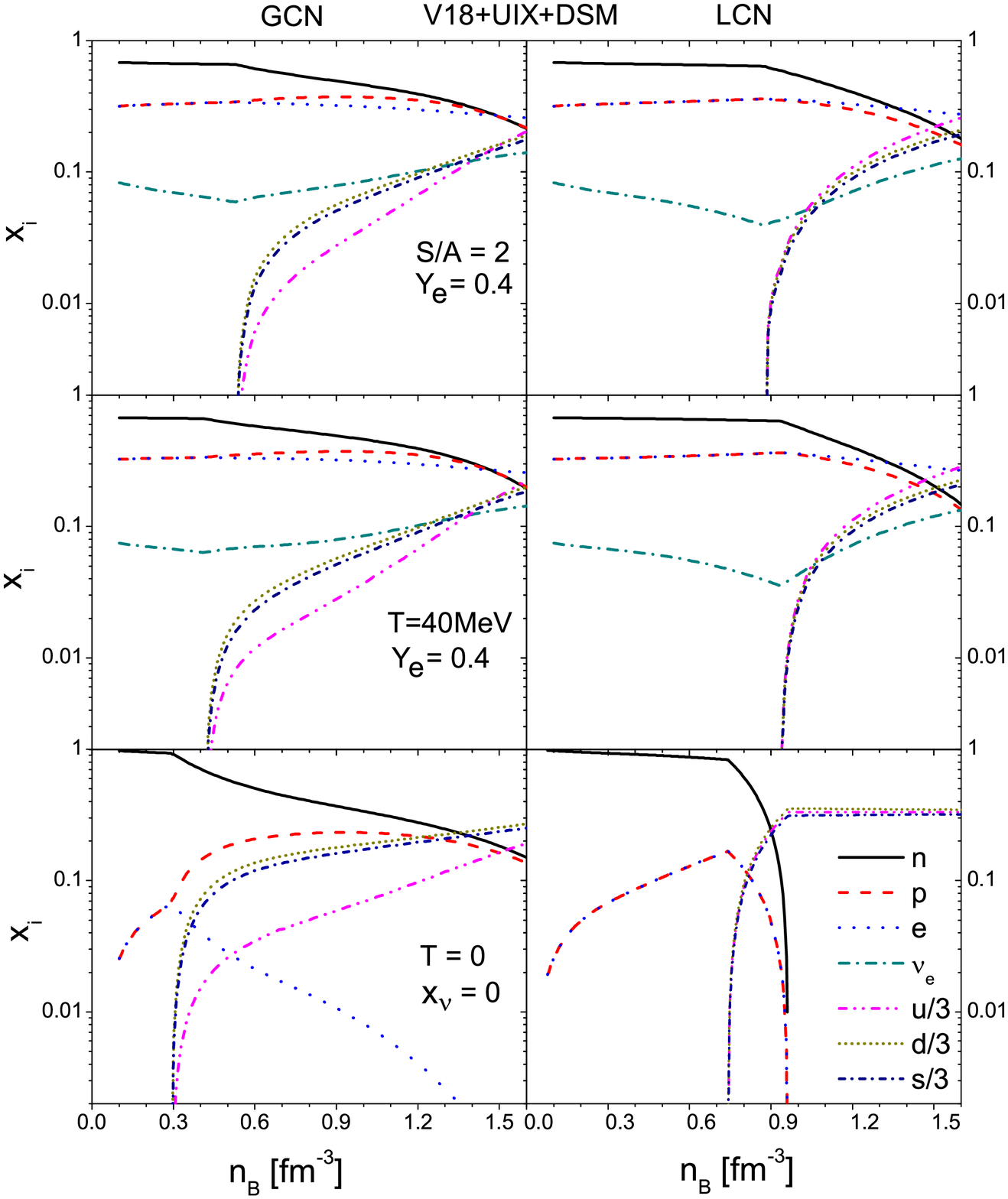}
\caption{
Same as Fig.~\ref{f:xi_mit},
but with the Dyson-Schwinger model for the quark phase.}
\label{f:xi_dsm}
\end{figure}

The opposite limiting case (LCN)
corresponds to a WS cell that is large relative to the electromagnetic
Debye screening length, and a large surface tension.
In this situation the electric charges are well screened inside the cell
and both QM and HM are locally charge neutral nearly everywhere,
\be
 n_C^H = n_C^Q = 0 \:,
\ee
except on a small boundary layer near the HQ interface, 
where a positively charged layer of HM and a negatively charged one of QM
are present and create a strong but very localized
electric field \citep{vos1,vos2}.
Consequently there occurs a sharp rise
$\delta\mu_C$
of the Coulomb potential at the HQ interface
and Eq.~(\ref{e:muc}) is modified to 
\be
 \mu_C^{H} = \mu_C^{Q} + \delta\mu_C \:, 
\ee
such that for example the electron density is now different in
hadron and quark phases.

In beta-stable untrapped matter this situation corresponds exactly
to the usual Maxwell construction, 
joining two charge-neutral phases by equality of pressure and baryon chemical
potential,
that is often employed for simplicity.
Including neutrino trapping with microscopic finite-size structures requires
always homogeneous neutrino densities, Eq.~(\ref{e:nnu}),
and therefore,
due to the unequal electron densities, 
in this case the trapping condition becomes a global one,
as expressed by Eq.~(\ref{e:nl}).
Due to the additional degree of freedom represented by the neutrino density,
then the LCN construction is realized in the PNS as an extended mixed phase 
involving a HQ coexistence region with a 
continuously varying pressure \citep{hempel2,hempel1,hempel3,yasurev}.

We have explained that the GCN and LCN constructions are in fact 
idealized scenarios 
that correspond to two opposite extremes of the microscopic treatment.
It has been pointed out in \citet{yasulet,yasurev}
that actually the LCN construction is closer to the full microscopic treatment
of finite-size effects than the GCN, 
and it is therefore of interest to compare the predictions of the two
constructions for the internal composition and other properties of PNS,
which we will do now.

\section{Results}
\label{s:res}

\subsection{Internal composition}

The relative particle populations are shown as a function of the baryon density 
in Fig.~\ref{f:xi_mit} for the bag model and in
Fig.~\ref{f:xi_dsm} for the DSM, 
for trapped matter with $Y_e=0.4$ and
i) entropy per baryon $S/A=2$ (upper panels), 
ii) temperature $T=40$ MeV (middle panels), and 
iii) untrapped and cold neutron star matter (lower panels).
The GCN (left panels) and LCN (right panels) calculations are compared.

There are big differences between the MIT and DSM regarding the HQ mixed phase
that have been pointed out in \citet{ourdsm1,ourdsm2}:
With the MIT model the HQ phase transition starts at fairly low baryon density
and a pure quark phase is reached at not too large density,
whereas with the DSM the onset of the mixed phase occurs at higher density
and the system remains in the mixed phase even at very large density.
Furthermore, hyperons are allowed with the MIT model,
where they might appear only in small fractions at low density
and are replaced by strange QM at higher density
(this can be seen in the central panels of Fig.~\ref{f:xi_mit}),
whereas they prevent any transition to QM with the DSM
and have to be excluded by hand in that case.
Therefore, these two very different quark models might be good candidates
to reveal important differences between the phase transition constructions
that we are examining.

In fact, comparing now the results obtained with both prescriptions
(left and right panels),
we observe behavior in line with the general properties mentioned before.
We remind that for cold NS matter (bottom panels) the LCN corresponds
to the usual Maxwell construction and the GCN to the bulk Gibbs construction;
and it is well known that the density range of the mixed phase
with the Gibbs construction is wider than the one with the Maxwell construction,
which can be seen in the plots.
This behavior remains also in trapped hot matter, where the GCN spans always
a wider density range than the LCN (with the MIT)
or begins at lower density (with the DSM).
In all cases the trapping condition shifts the onset of the mixed phase 
to slightly higher density.

The differences between the LCN and GCN constructions are fairly small
for the MIT model, but significant for the DSM:
Here the GCN (Bulk Gibbs) mixed phase occurs in a much wider density interval
than the LCN (Maxwell) one.
Apart from the Maxwell construction for cold NSs, the matter remains in the
mixed phase and pure QM is never reached.
This variance is due to the qualitatively different density dependence
of the effective bag constant in the MIT and DSM,
see \citet{ourdsm1,ourdsm2}.

\begin{figure}[t]
\includegraphics[scale=0.35,bb=50 50 814 820]{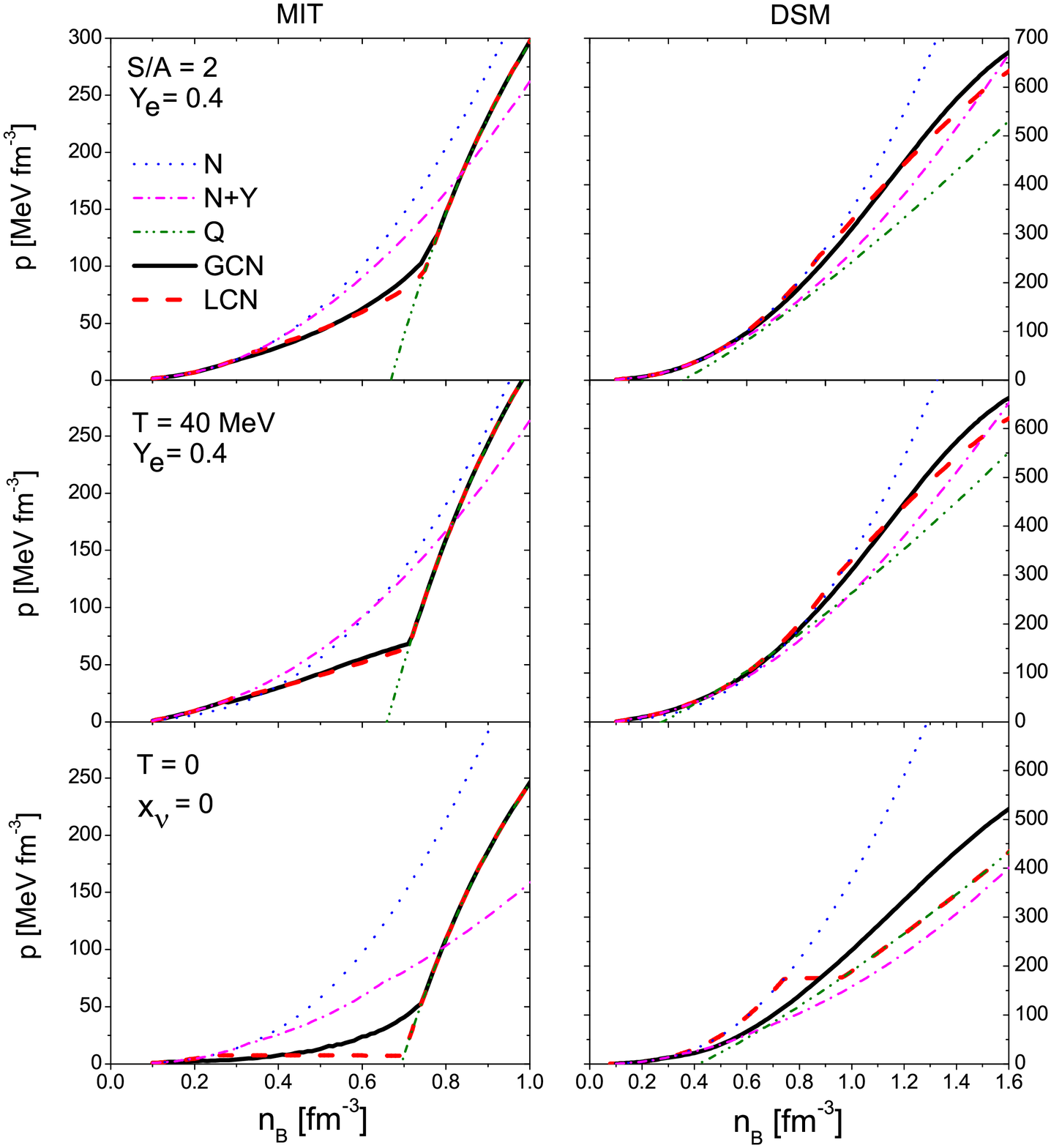}
\caption{
The pressure vs.~baryon density 
for the same stellar configurations as in Fig.~\ref{f:xi_mit},
and obtained with the MIT (left panels)
and the DSM (right panels) quark models.}
\label{f:eos}
\end{figure}

\subsection{Equation of state}

In Fig.~\ref{f:eos} the EOS $p(n_B)$ is displayed
for the different stellar configurations, quark models,
and mixed phase constructions as before.
For comparison also the pure phases 
(nucleons only, nucleons+hyperons, quarks) 
are shown.

We observe only minor differences between GCN and LCN
in the hot and trapped matter,
even for the DSM, where the particle fractions are quite different
in both cases.
Stronger differences between GCN and LCN appear in the cold case, where
a plateau in the pressure shows up for the LCN (Maxwell) calculation.

Thus if during the temporal evolution the system would remain in the
LCN phase, the extended mixed phase region would gradually disappear.
However, one should remember that both LCN and GCN are highly idealized 
constructions and the true behavior depends on currently uncertain microphysics
like the HQ surface tension.

\subsection{Stellar structure}

\begin{figure}[t]
\includegraphics[scale=0.54,bb=30 90 794 850]{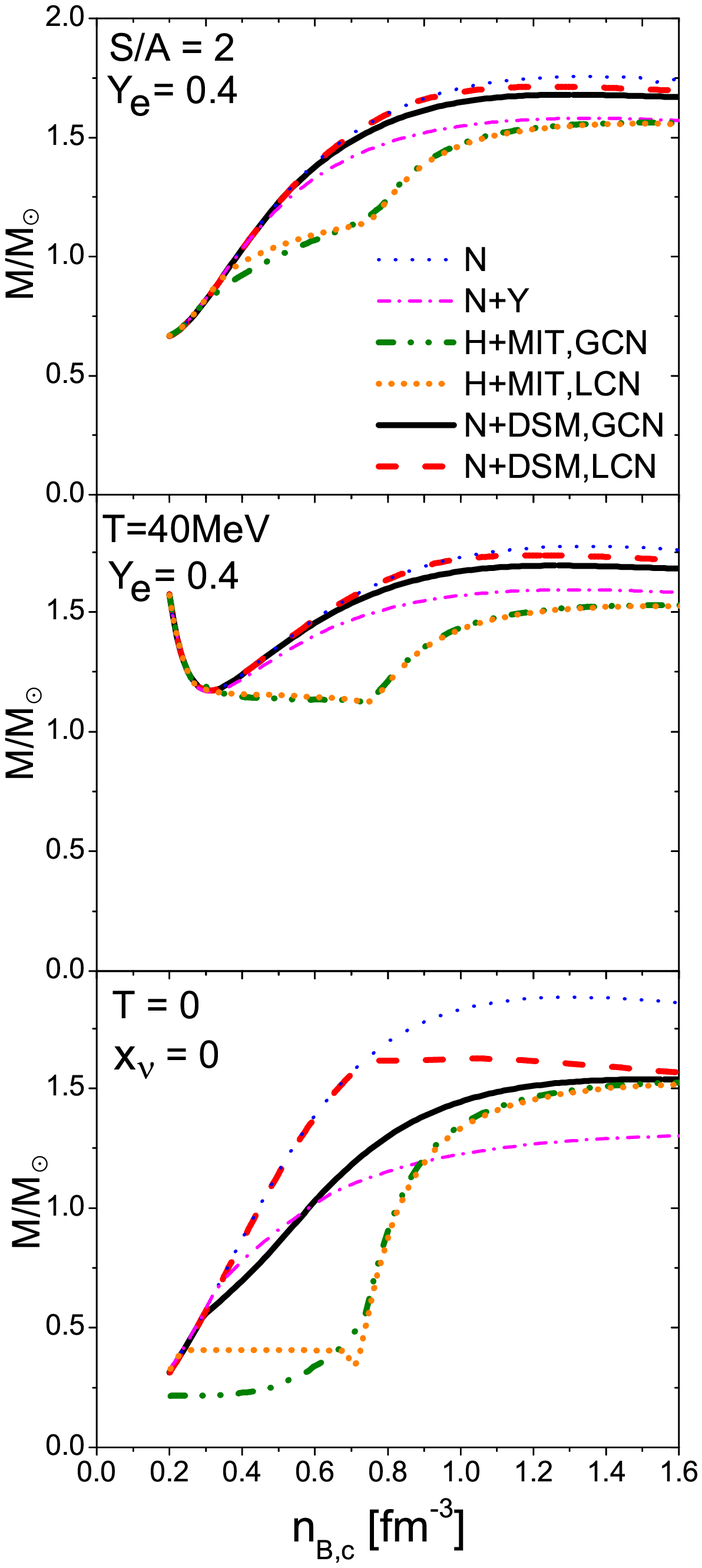}
\caption{
The gravitational mass vs.~central baryon density
for different stellar configurations and EOS.}
\label{f:mrho}
\end{figure}

Once the EOS is known, the stable configurations of a (P)NS can be obtained
from the well-known hydrostatic equilibrium equations
of Tolman, Oppenheimer, and Volkov \citep{shapiro}.
In the low-density range, where nucleonic clustering sets in,
we cannot use the BHF approach,
and therefore we join \citep{isen2} the BHF EOS to the
finite-temperature or -entropy EOS of \citet{shen1,shen2},
which is more appropriate at densities below
$n_B \lesssim 0.07\;{\rm fm}^{-3}$,
since it does include the treatment of finite nuclei.

Our results for the gravitational mass as a function of the
central baryon density, 
for the different stellar configurations and 
using the different EOS introduced previously,
are displayed in Fig.~\ref{f:mrho}.
We observe in NS matter the strong softening effect
of the hyperons (dash-dotted purple curves)
on the purely nucleonic configurations (dotted blue curves),  
which is however strongly reduced in trapped matter,
because the hyperon concentrations remain smaller \citep{rep,pnsy}.

In the case of the MIT model, the mass-density relations
obtained with the GCN or LCN constructions 
(dash-dot-dotted green vs short-dashed orange curves)
are nearly indistinguishable,
apart from the unphysical low-mass region, where the Maxwell construction
can be recognized in the NS configuration with LCN.
The maximum mass of the hybrid stars decreases slightly with respect to
both the nucleonic and the hyperonic stars in the trapped cases,
whereas it increases (decreases) with respect to cold hyperon (nucleon) NS.
In NS this is due to the fact that the hyperon population is suppressed
by the onset of quarks,
whereas in PNS the trapping reduces the hyperon population.
In all cases, with the HQ phase transition,
the value of the maximum mass is about 1.5 $M_\odot$
and thus rather low,
as is a general feature of the MIT model \citep{alford}.

If the DSM is used, 
the differences between GCN and LCN 
(solid black vs dashed red curves)
are slightly larger,
in particular for NS, where the LCN leads to unstable configurations
at the onset of the quark phase.
Nevertheless the differences between GCN and LCN maximum masses 
are insignificant in all configurations.
In this case the phase transition takes place only
if hyperons are excluded from the hadronic phase,
and the value of the maximum mass for the hybrid configuration always
decreases with respect to the purely nucleonic star.
In particular, for PNS the maximum mass is about $1.75\;M_\odot$,
while for NS it is slightly smaller.
 
These values for the maximum mass of hybrid (P)NS depend on the 
choice of the nucleonic three-body force and on the details of the DSM.
Larger values could eventually be reached with different parameter
choices \citep{ourdsm1,ourdsm2}, 
in agreement with the current observational data \citep{demo},
which is in contrast to the case with the MIT model. 
However, the exact value of the maximum mass of a (P)NS is still an open problem
[see also \citet{heavy}]
and not the purpose of the present article.

\section{Conclusions}
\label{s:end}

In this article we discussed the occurrence of a hadron-quark mixed phase
in the interior of hybrid (proto)neutron stars.
We explained the physical origin and justification of the idealized LCN and GCN
phase transition constructions, which represent two opposite limiting
cases of the microscopic treatment of electromagnetic finite-size effects,
and examined their consequences with two very different quark models.

While indeed the internal composition of hybrid (proto)stars turns out to be
very different with both constructions, the impact on the equation
of state and masses is very much reduced, so that these global observables
could hardly serve as an indication for the type of phase transition
and thus the internal stellar structure.

Therefore, for a true understanding of the nature of the mixed phase, detailed 
microscopic investigations of the finite size effects and their 
importance for the stellar microphysics (cooling, transport, oscillations, ...)
are required.

\section*{Acknowledgments}

We acknowledge useful discussions with T.~Maruyama and T.~Tatsumi.
This work was partially supported by CompStar,
a Research Networking Programme of the European Science Foundation,
and by the MIUR-PRIN Project \hbox{No.~2008KRBZTR}.

\bibliographystyle{aa}  
\bibliography{mixpns}   
\end{document}